\documentclass[runningheads,a4paper]{llncs}

\usepackage{graphicx}
\usepackage{latexsym}
\usepackage{amsmath}
\usepackage{amssymb}
\usepackage{fancyvrb}
\usepackage{subfigure}
\usepackage{colortbl}
\usepackage{enumerate}
\usepackage{pifont}
\usepackage{stmaryrd}
\usepackage{textcomp}
\usepackage{fncylab}
\usepackage{multirow}
\usepackage{paralist}
\usepackage{wrapfig}
\usepackage{colortbl}
\usepackage{longtable}
\usepackage{lscape}
\usepackage{makeidx}

\graphicspath {{./} {Figures/}}

\newcolumntype{H}{>{\columncolor{black}\color{white}}c}

\graphicspath {{./} {Figures/}}

\newcommand{\keywords}[1]{\par\addvspace\baselineskip
\noindent\keywordname\enspace\ignorespaces#1}

\begin{document}
\title{A Query Language for Summarizing and Analyzing Business Process Data}

\titlerunning{BP-SPARQL}

\author{Amin Beheshti\inst{1} \and Boualem Benatallah\inst{2} \and Hamid Reza Motahari-Nezhad\inst{3} \and \\
Samira Ghodratnama\inst{1} \and Farhad Amouzgar\inst{1}}
\institute{Macquarie University, Sydney, Australia\\
\email{amin.beheshti@mq.edu.au} \\
\email{\{samira.ghodratnama,farhad.amouzgar\}@hdr.mq.edu.au}\\
\and University of New South Wales, Sydney, Australia\\  \email{boualem@cse.unsw.edu.au}\\
\and EY AI Lab, USA\\ \email{hamid.motahari@ey.com}
}

\authorrunning{A. Beheshti, et al.}

\maketitle

\begin{abstract}
In modern enterprises, Business Processes (BPs) are realized over a mix of workflows, IT systems, Web services and direct collaborations of people.
Accordingly, process data (i.e., BP execution data such as logs containing events, interaction messages and other process artifacts) is scattered across several systems and data sources, and increasingly show all typical properties of the Big Data.
Understanding the execution of process data is challenging as
key business insights remain hidden in the interactions among process entities: most objects are interconnected, forming complex, heterogeneous but often semi-structured networks.
In the context of business processes, we consider the Big Data problem as a massive number of interconnected data islands from personal, shared and business data.
We present a framework to model process data as graphs, i.e., Process Graph,
and present abstractions to summarize the process graph and to discover concept hierarchies for entities based on both data objects and their interactions in process graphs.
We present a language, namely BP-SPARQL, for the explorative querying and understanding of process graphs from various user perspectives.
We have implemented a scalable architecture for querying, exploration and analysis of process graphs.
We report on experiments performed on both synthetic and real-world datasets that show the viability and efficiency of the approach.

\keywords{Business Processes, Query Processing, Big Process Data, Process Data Science}
\end{abstract}

\section{Introduction}
\label{intro}

A business process is a set of coordinated tasks and activities, carried out manually or automatically, to achieve a business objective or goal~\cite{DBLP:conf/bpm/AalstHW03}.
In modern enterprises, Business Processes (BPs) are realized over a mix of workflows, IT systems, Web services and direct collaborations of people.
In such information systems, business process analysis over a wide range of systems, services and softwares (that implement the actual business processes of enterprises) is required.
This is challenging as in today's knowledge-, service-, and cloud-based economy the information about process execution is scattered across several systems and data sources.
Consequently, process logs increasingly come to show all typical properties of the Big Data~\cite{mcafee2012big}: wide physical distribution, diversity of formats, non-standard data models, independently-managed and heterogeneous semantics. We use the term \emph{process data} to refer to such large hybrid collections of heterogeneous and partially unstructured process related data.

Understanding process data requires scalable and process-aware methods to support querying, exploration and analysis of the process data in the enterprise because:
(i)~with the large volume of data and their constant growth, the process data analysis and querying method should be able to scale well; 
and
(ii)~the process data analysis and querying method should enable users to express there needs using process-level abstractions.
Besides the need to support process-level abstractions in process data analysis scenarios, the other challenge is the need for scalable analysis techniques to support Big Data analysis. Similar to Big Data processing platforms~\cite{zikopoulos2011understanding}, such analysis and querying methods should offer automatic parallelization and distribution of large-scale computations, combined with techniques that achieve high performance on large clusters of commodity PCs, e.g., cloud-based infrastructure, and be designed to meet the challenges of process data representation that capture the relationships among data.

In this chapter, we present a summary of our previous work~\cite{BPM11,caise13,POLAP,ProcessAtlas} in organizing, querying, and analyzing business processes' data.
We introduce BP-SPARQL, a query language for summarizing and analyzing process data.
BP-SPARQL supports a graph-based representation of data relationships~\cite{DREAM,DBLP:journals/cluster/BatarfiSFNBBS15}, and enables exploring, analyzing and querying process data and their relationships by superimposing process abstractions over an entity relationship graph, formed over entities in process-related repositories.
This will provide analysts with process related entities (such as process event, business artifacts, and actors), abstractions (such as case, process instances graph, and process model) and functions (such as correlation condition discovery, regular expressions, and process discovery algorithms) as first class concepts and operations.
BP-SPARQL supports various querying needs such as:
entity-level (artifacts, events and activities),
summarization (OLAP Style, Group Style and Partition Style),
relationship (Regular-Expression, Path Condition and Path Node),
meta-data (Time and Provenance) and
user-defined queries.

The remainder of this chapter is organized as follows.
In Section~\ref{bk}, we present the background and contributions overview. We introduce the process graph model in Section~\ref{dataModel}.
Section~\ref{sum} presents the abstractions used to summarize the process data.
In Section~\ref{queryFT}, we present the query framework and in Section~\ref{MapReduceSupport} we present MapReduce techniques to scale the analysis.
In Section~\ref{imp}, we describe the implementation and the evaluation experiments.
Finally, we position our approach within the Process Querying Framework
in Section~\ref{RelWork}, before concluding the chapter in Section~\ref{Conclusion}.

\section{Background and Contributions Overview}
\label{bk}

The problem of understanding the behavior of information systems as well as the processes and services they support has become a priority in medium and large enterprises. This is demonstrated by the proliferation of tools for the analysis of process executions, system interactions, and system dependencies, and by recent research work in process data warehousing and process
discovery. Indeed, the adoption of business process intelligence techniques for process improvement is the primary concern for medium and large companies.
In this context, identifying business needs and determining solutions to business problems requires the analysis of business process data:
this enables discovering useful information, suggesting conclusions, and supporting decision making for enterprises.

In order to understand available process data (events, business artifacts, data records in databases, etc.) in the context of process execution, we need to represent them, understand their relationships and enable the analysis of those relationships from the process execution perspective.
To achieve this, it is possible to represent process-related data as entities and any relationships among them (e.g., events relationships in process logs with artifacts, etc.) in entity-relationship graphs.
In this context, business analytics can facilitate the analysis of process data in a detailed and intelligent way through describing the applications of analysis, data, and systematic reasoning~\cite{POLAP}. Consequently, an analyst can gather more complete insights using data-driven techniques such as modeling, summarization and filtering.

\textbf{Motivating Scenario.}
Modern business processes (BPs) are rarely supported by a single, centralized workflow engine.
Instead, BPs are realized using a number of autonomous systems, Web services, and collaboration of people. As an example, consider the banking industry scenario. Recently, there is a movement happening in the banking industry to modernize core systems for providing solutions to account management, deposits, loans, credit cards, and the like. The goal is to provide flexibility to quickly and efficiently respond to new business requirements.
In order to understand the process analysis challenges, let's consider a real world case in the loan scenario, where Adam (a customer) plans to buy a property. He needs to find a lending bank. He can use various crowdsourcing services~\cite{CrowdCorrect} (e.g., Amazon Mechanical Turk\footnote{https://www.mturk.com/}) or visit mortgage bank to find a proper bank. Then, he needs to contact the bank through one of many channels and start the loan pre-approval process. After that he needs to visit various Websites or real-estate services to find a property. Meanwhile he can use social Websites (e.g., Facebook or Twitter) to socialize the problem of buying a property and ask for others' opinions to find a proper suburb. After finding a property, he needs to choose a solicitor to start the process of buying the property. Lots of other process can be executed in between. For example, the bank may outsource the process of evaluating the property or analyzing Adam's income to other companies.

In this scenario, the data relevant to the business process of bank is scattered across multiple systems and in many situations, stakeholders can be aware of processes but they are not able to track or understand them:  it is important to maintain the vital signs of customers by analyzing the process data. This task is challenging as:
(i)~a massive number of interconnected data islands (from personal, shared and business data) need to be processed;
(ii)~processing this data requires scalable methods; and
(iii)~analyzing this data depends on the perspective of the process analyst, e.g.,
``\emph{Where is loan-order $\#756$? What happened to it? Is it blocked?
Give me all the documents and information related to the processing of loan-order $\#756$?
What is the typical path of loan orders? What is the process flow for it?
What are the dependencies between loan applications $A_1$ and $A_2$?
How much time and resources are spent in processing loan orders that are eventually rejected?
Can I replace X (i.e., a service or a person) by Y?
Where data came from? How it was generated? Who was involved in processing file X?
At which stage do loan orders get rejected? How many loan orders are rejected in the time period between $\tau_1$ and $\tau_2$?
What is the average time spent on processing loan orders? Where do delays occur? Etc.}".

Answering these questions in today's knowledge-, service-, and cloud-based economy is challenging as the information about process execution is scattered across several systems and data sources. Therefore, businesses need to manage unstructured, data-intensive and knowledge-driven processes rather than well predefined business processes. Under such conditions, organizing, querying and analyzing  process data becomes of a great practical value but clearly a very challenging task as well.
In the following, we provide an overview of the main contributions of this chapter:

1) \emph{Process-aware abstractions for querying and representing process data}~\cite{BPM11,caise13}:
We introduce a graph-based model to represent all process-related data as entities and relationships among those entities in an entity-relationship graph (ER Graph).
To enable analyzing process entity relationship graph directly, still in a process-aware context, we define two main abstractions for summarizing the process data (modeled as an ER graph):
\textit{Folder} nodes (a container for the results of a query that returns a collection of related process entities) and \textit{path} nodes (a container for the results of a query that returns a collection of related paths found in the entity-relationship graph). We present other extensions of folder and path nodes including \emph{timed} folder nodes, and also process metadata queries to support process analysis needs in contexts such as process entity provenance~\cite{DBLP:journals/corr/abs-1211-5009} and artifact versioning. These summarization abstractions and functions offer a comprehensive set that covers most prominent needs of querying process-related data from different systems and services.

2) \emph{Summarizing process data}~\cite{POLAP}:
We introduce a framework and a set of methods to support scalable graph-based OLAP (Online analytical processing) queries over process execution data.
The goal is to facilitate the analytics over the ER graph through summarizing the process graph and providing multiple views at different granularities.
To achieve this goal, we present a model for process OLAP (P-OLAP) and define OLAP specific abstractions in process context such as process cubes, dimensions, and cells. We present a MapReduce-based graph processing engine, to support Big Data analytics over process data.
We identify useful machine learning algorithms~\cite{iSheets} and provide an external algorithm controller to enable summarizing the process data (modeled as an ER graph), by extracting complex data structures such as timeseries, hierarchies, patterns and subgraphs. 
We define a set of domain-specific abstractions and functions such as process event, process instance, events correlation condition, process discovery algorithm, correlation condition discovery algorithm, and regular expression to summarize the process data and to enable the querying and analysis of relationships among process-related entities.

3) \emph{Scalable process data analysis and querying methods}~\cite{ProcessAtlas}.
In order to support the scalable exploration and analysis of process data, 
we present a domain specific language, namely BP-SPARQL, that supports the querying of process data (modeled as an ER graph) using the above mentioned process-level abstractions. BP-SPARQL translates process-level queries into graph-level abstractions and queries. To support the scalable and efficient analysis over process data, BP-SPARQL is implemented over the MapReduce\footnote{The popular MapReduce~\cite{MapReduce} scalable data processing framework, and its opensource realization Hadoop~\cite{Hadoop}, offer a scalable data flow programming model.} framework by providing a data mapping layer for the automatic translation of queries into MapReduce operations. For this purpose, we designed and implemented a translation from BP-SPARQL to Hadoop PigLatin~\cite{olston2008pig}, where the resulting PigLatin program is translated into a sequence of MapReduce operations and executed in parallel on a Hadoop cluster. The proposed translation offers an easy and efficient way to take advantage of the performance and scalability of Hadoop for the distributed and parallelized execution of BP-SPARQL queries on large graph datasets.
BP-SPARQL supports the following query types:
(i)~entity level queries: for querying process related entities, e.g., business artifacts, actors, and activities;
(ii)~relationship queries: for discovering relationships and patterns among process entities using regular expressions;
(iii)~summarization queries: these queries allows for analyzing case-based processes to find potential process instances by supporting OLAP Style, Group Style and Partition Style queries;
(iv)~metadata queries: for analyzing the evolution of business artifacts and their provenance over time; and
(v)~user-defined queries.

\section{Process Abstractions}
\label{dataModel}

\begin{figure}
\centering
  \includegraphics[scale=0.95]{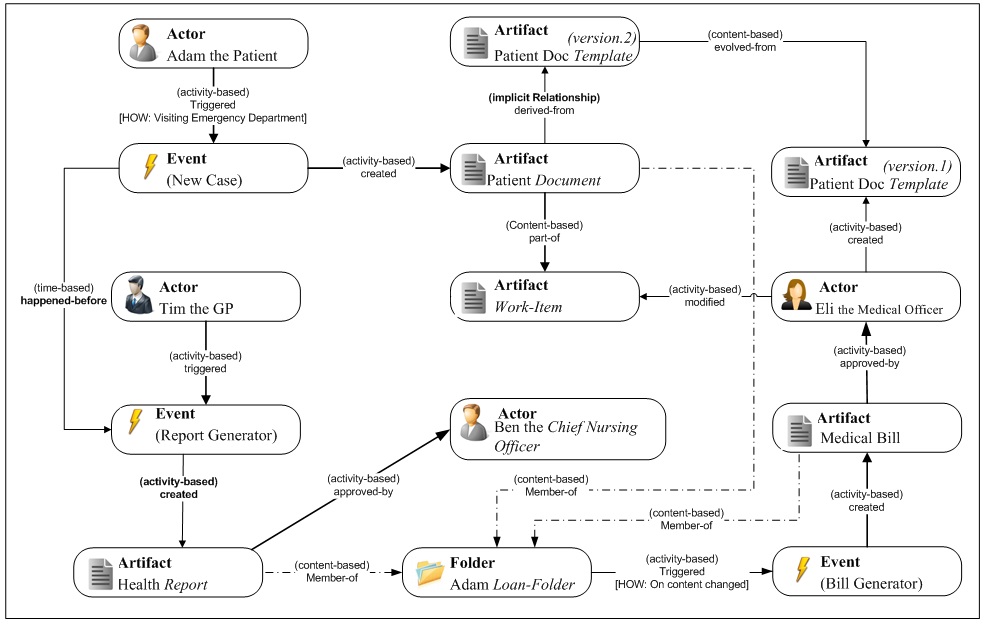}
  \caption{A sample graph of relationships among entities in the motivating banking scenario presented in Section~\ref{bk}.}\label{canonical-graph}
\end{figure}

In this section, we provide a summary of our previous work~\cite{BPM11,caise13,DBLP:conf/wise/BeheshtiBNA12} on modeling process data as an ER Graph. 
To organize the process data, we introduce a graph-based data model.
The data model includes entities (e.g., events, artifacts, and actors), their relationships, and abstractions which act as higher level entities to store and browse the results of queries for follow-on analysis. The data model is based on the RDF~\cite{SPARQL} data representation. 

\begin{definition}\label{definition:Entity} \emph{
(Entity) An entity $N$ is a data object that exists separately and has a unique identity.
}\end{definition}

Entities can be structured (e.g., customer, bank, and branch) or unstructured (body of an e-mail message).
Structured entities are instances of entity types. This entity model offers flexibility when types are unknown and takes advantage of structure when types are known. Specific types of entities include:

\begin{itemize}
  \item (Entity: Business Artifact), is a digital representation of something, i.e., data object, that exists separately as a single and complete unit and has a unique identity. A business artifact is a \emph{mutable} object, i.e., its attributes (and their values) are able and are likely to change over periods of time. An artifact $A$ is represented by a set of attributes $\{a_1, a_2, ..., a_k\}$, where $k$ represents the number of attributes. An artifact may appear in many versions. A \emph{version} $v$ of a business artifact is an \emph{immutable} copy of the artifact at a certain point in time. A business artifact $A$ can be represented by a set of versions $\{v_1, v_2, ..., v_n\}$, where $n$ represents the number of versions. A business artifact can capture its current state as a version and can restore its state by loading it. Each version is represented as a data object that exists separately and has a unique identity. Each version $v_i$ consists of a snapshot, a list of its parent versions, and meta-data, such as commit message, author, owner, or time of creation. In order to represent the history of a business artifact, it is important to create archives containing all previous states of an artifact. The archive allows us to easily answer certain temporal queries such as retrieval of any specific version and finding the history of an artifact.
  \item (Entity: Actor), An actor $R$ is an entity acting as a catalyst of an activity, e.g., a person or a piece of software that acts for a user or other programs. A process may have more than one actor enabling, facilitating, controlling, affecting its execution.
  \item (Entity: Event) An event is an object representing an activity performed by an actor. An event $E$ can be presented as the set $\{R,\tau,D\}$, where $R$ is an actor (i.e., a person or device) executing or initiating an activity, $\tau$ is a timestamp of the event, and $D$ is a set of data elements recorded with the event (e.g., the size of an order). We assume that each distinct event does not have a temporal duration. For instance, an event may indicate an arrival of a loan request (as an XML document) from a bank branch, an arrival of a credit card purchase order from a business partner, or a completion of a transmission or a transaction.
\end{itemize}

\begin{definition}\label{definition:Relationship} \emph{
(Relationship) A \emph{relationship} is a directed link between a pair of entities, which is associated with a predicate defined on the attributes of entities that characterizes the relationship. A relationship can be \emph{explicit}, such as `was triggered by' in \textcolor{black}{$event_1$ $\xrightarrow{(\emph{wasTriggeredBy})}$ $event_2$}
in business processes (BPs) execution log, or \emph{implicit}, such as a relationship between an entity and a larger (composite) entity that can be inferred from the nodes.
An entity is related to other entities by time (time-based), content (content-based), or activity (activity-based):
}\end{definition}

\begin{itemize}
  \item (Time-based Relationships) Time is the relationship that orders events, for example, $event_A$ happened before $event_B$. A timestamp is attached to an event where every timestamp value is unique and accurately represents an instant in time. Considering activities $A_{\tau1}$ and $A_{\tau2}$, where $\tau$ is a timestamp of an event, $A_{\tau1}$ happened before $A_{\tau2}$ if and only if $\tau1 < \tau2$.
  \item (Content-based Relationships) When talking about content, we refer to entity attributes. In process context, we consider content-based relationships as correlation condition-based relationships, where a \emph{correlation condition}~\cite{hamidEventCorrelation} is a binary predicate defined over attributes of two entities $E_x$ and $E_y$ and denoted by $\psi(E_x,E_y)$. This predicate is true when $E_x$ and $E_y$ are correlated and false otherwise. A correlation condition $\psi$ allows to partition an entity-relationship graph into sets of related entities. 
  \item (Activity-based Relationships) This is a type of relationship between two entities that is established as the result of performing an activity. In this context, an activity can be described by a set of attributes such as: (i)~What (types of activity), (ii)~How (actions such as creation, transformation, or use), (iii)~When (the timestamp in which the activity has occurred), (iv)~Who (an actor that enables, facilitates, controls, or affects the activity execution), (v)~Where (the organization/department where the activity happened), (vi) Which (the system which hosts the activity), and (vii)~Why (the goal behind the activity, e.g., fulfilment of a specific phase).
\end{itemize}

Other types of relationships are discussed in~\cite{DBLP:conf/fase/BarrosDDW07}.
Figure~\ref{canonical-graph} illustrates a sample process data modeled as an ER graph, illustrating possible relationships among entities in the motivating (banking) scenario.
Next, we define RDF Triples as a representation of relationships.

\begin{definition}\label{definition:rdfTriple} \emph{
(RDF Triple)
The RDF terminology $T$ is defined as the union of three pairwise disjoint infinite sets of terms: the set $U$ of URI references,
the set $L$ of literals, and the set $B$ of blanks. The set $U \cup L$ of names is called the vocabulary.
An RDF triple (\textbf{s}ubject, \textbf{p}redicate, \textbf{o}bject) is an element of $(s,p,o) \in (U \cup B) \times U \times T$, where $s$ is a subject, $p$ is a predicate and $o$ is an object.
}\end{definition}

An RDF graph is a finite set of RDF triples. An RDF triple can be viewed as a relationship (an arc) from subject $s$ to object $o$, where predicate $p$ is used to label the relationship. This is represented as \textcolor{black}{s $\xrightarrow{(p)}$ o}.
Next, we define the Entity-Relationship (ER) Graph which is capable of representing process data as entities and relationships among those entities. 
An ER graph may contain all possible relationships (from Definition~2) among its nodes. 

\begin{definition}\label{definition:erGraph}
\emph{
An Entity-Relationship Graph $G=(V, E)$ is a directed graph with no directed cycles, where $V$ is a set of nodes and $E \subseteq (V \times V )$ is a set of ordered pairs called edges.
The Entity-Relationship (ER) graph $G$ is modeled using the RDF data model to make statements about resources (in expressions of the form subject-predicate-object, known as triples), where a resource in an ER graph defined as follows:
(i)~The sets $V_G$ and $E_G$ are resources; and (ii)~The set of ER graphs is closed under intersection, union and set difference: let $G_1$ and $G_2$ be two ER graphs, then $G_1 \cup G_2$, $G_1 \cap G_2$, and $G_1 - G_2$ are ER graphs.
}\end{definition}

\section{Summarizing Big Process Data}
\label{sum}

In this section, we provide a summary of our previous work~\cite{POLAP,DBLP:conf/wise/BeheshtiBNA12} on summarizing the process data.
We introduce a framework and a set of methods to support scalable graph-based OLAP (On-Line Analytical Processing) analytics over process execution data.



We present two types of queries to summarize the process data (modeled as ER graphs), namely ``Correlation Condition'' and ``Regular Expression'':


\begin{itemize}
  \item \emph{\textbf{Correlation Condition}} is a binary predicate defined on the attributes of events that allows to identify whether two or more events are potentially related to the same execution instance of a process~\cite{BPM11}. In particular, a correlation condition takes ER graph and a predicate (as input), applies an algorithm (e.g., the one introduced in our previous work~\cite{hamidEventCorrelation}) to find the set of correlated events for that condition, and returns a set of ER subgraphs $\{G'_1,G'_2,...,G'_k\}$ representing correlated events. Formally, $G' = (V', E')$ is an ER subgraph of an ER graph G=(V, E) iff $V' \subseteq V$ and $E' \subseteq E$ and $((v_1,v_2)\in E' \rightarrow \{v_1,v_2\} \subseteq V')$.
  \item \emph{\textbf{Regular Expressions}} is a function which represents a relation between ER graph $G$ and a set of paths $\{P_1,P_2,...,P_m\}$, where a \emph{path} $P$ is a transitive relationship between entities capturing sequences of edges from the start entity to the end entity. A path can be represented as a sequence of RDF triples, where the object of each triple in the sequence coincides with the subject of its immediate successor. 
      We developed a regular expression processor which supports optional elements, loops, alternation and grouping~\cite{DBLP:conf/edbt/BeheshtiBM16}.
\end{itemize}

\begin{figure}[t]
\centering
  \includegraphics[scale=0.35]{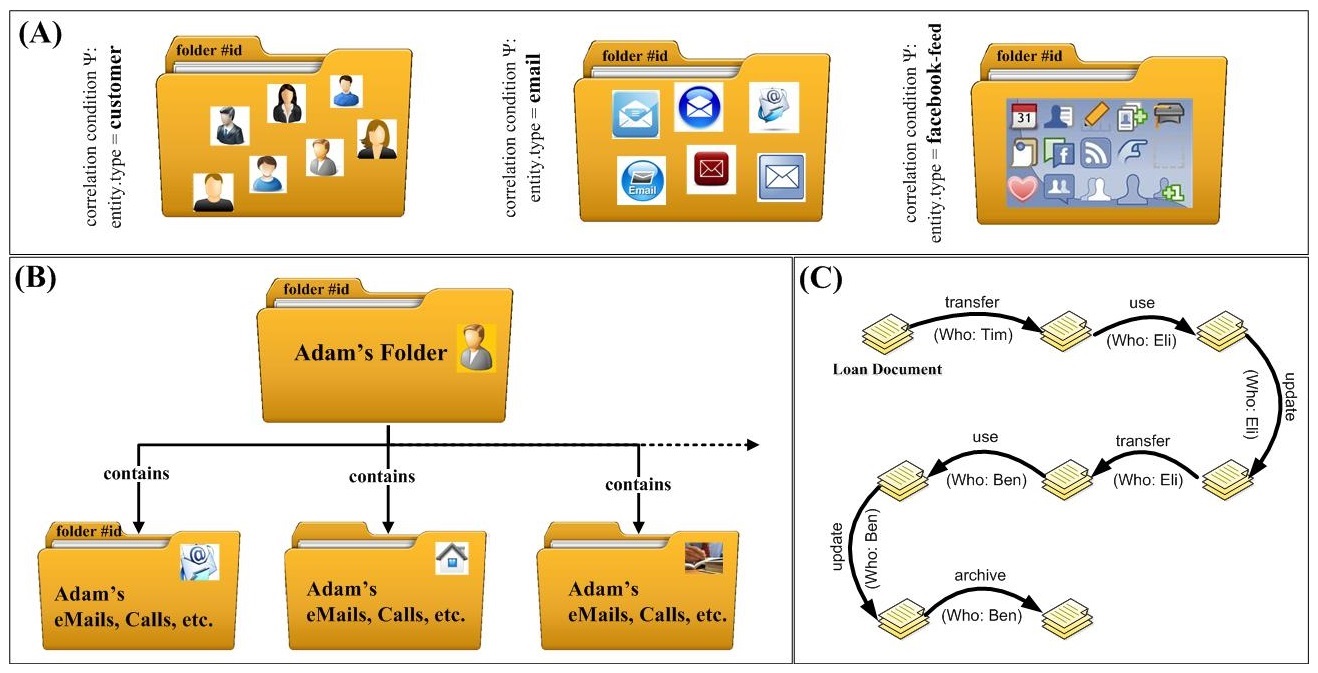}
  \caption{An example of a basic folder (A), a high-level folder (B), and a path (C).}\label{example-folder}
\end{figure}

An example path is illustrated in Fig.~\ref{example-folder}(C). This path depicts that the loan document was transferred by Tim, used and transferred by Eli to Ben, who updated the document and archived it.
Paths (defined by regular expressions) are written by domain experts and BP-SPARQL takes care of the processing and optimization that is needed for efficient crawling, analyzing and querying of the ER graph.
In particular, a regular expression $RE$ is an operator that specifies a search pattern and can result in a set of paths. Related paths can be stored in a path node~\cite{BPM11}, i.e., a container for 
a collection of related paths found in the entity-relationship graph.

\begin{definition}\label{definition:pathNode} \emph{
(Path Node~\cite{BPM11}) A path-node is a place holder for a set of related paths: these paths are the result of a given query that requires grouping graph patterns in a certain way.
We define a path node as a triple of $(V_{start},V_{end},RE)$ in which $V_{start}$ is the starting node, $V_{end}$ is the ending node, and $RE$ is a regular expression.
We use existing reachability approaches to verify whether an entity is reachable from another entity in the graph.
Path nodes can be timed.
A \emph{timed}-path node~\cite{caise13}, is defined as a timed container for a set of related entities which are connected through \emph{transitive} relationships, e.g., it is able to trace the evolution of patterns among entities over time as new entities and relationships are added over time.
}\end{definition}

Besides applying queries to the relationships in ER graphs (where the result will be a set of related paths and stored in path-nodes), queries can be applied to the content of the entities in the ER graph (without considering their relationships), i.e., the query may define criteria beyond existing relationships among entities to allow for discovering other relationships.
In this case, the result will be a set of correlated entities and possible relationships among them stored in \emph{folder nodes}.

\begin{definition}\label{definition:folder} \emph{
(Folder Node)
A folder-node~\cite{BPM11} contains a set of correlated entities that are the result of applying a function (e.g., Correlation Condition) on entity attributes.
The folder concept is akin to that of a database view, defined on a graph. However, a folder is part of the graph and creates a higher level node that other queries could be executed on. \emph{Basic folders} can be used to aggregate same entity types, e.g., related eMails, facebook feeds or activities. Fig.~\ref{example-folder}(A) illustrates a set of basic folders.
\emph{High-level folders}, illustrated in Fig.~\ref{example-folder}(B) can be used to create and query a set of related folders with relationships at higher levels of abstraction. A folder may have a set of attributes that describes it. Folder nodes can be timed.
Timed-folders~\cite{caise13}, document the evolution of folder node by adapting a monitoring code snippet. New members can be added to or removed from a timed folder over time.
}\end{definition}

Now, to support scalable graph-based OLAP analytics over process data, we define a mapping from process data 
into a graph model. 

\begin{definition}\label{definition:case} \emph{
(Process Instance) A Process Instance can be defined as a path $P$ in which the nodes in $P$ are of type `event', are in chronological order, and all the relationships in the path are of type activity-based relationships.
}\end{definition}

A collection of (disconnected) ER graphs, each representing a process instance, can be considered as a process instance graph and defined as follows:

\begin{definition}\label{definition:ProcessGraph} \emph{
(Process Instances Graph) A Process Instances Graph is a set of related process instances.
A Process Instances Graph is the result of a Correlation Condition query (stored in a folder node) or a Regular Expression query (stored in a path node).
%
}\end{definition}

\begin{landscape}
 \begin{figure} [t]
 \hspace*{-4.0cm}
 \centering
 \includegraphics[width=1.9\textwidth]{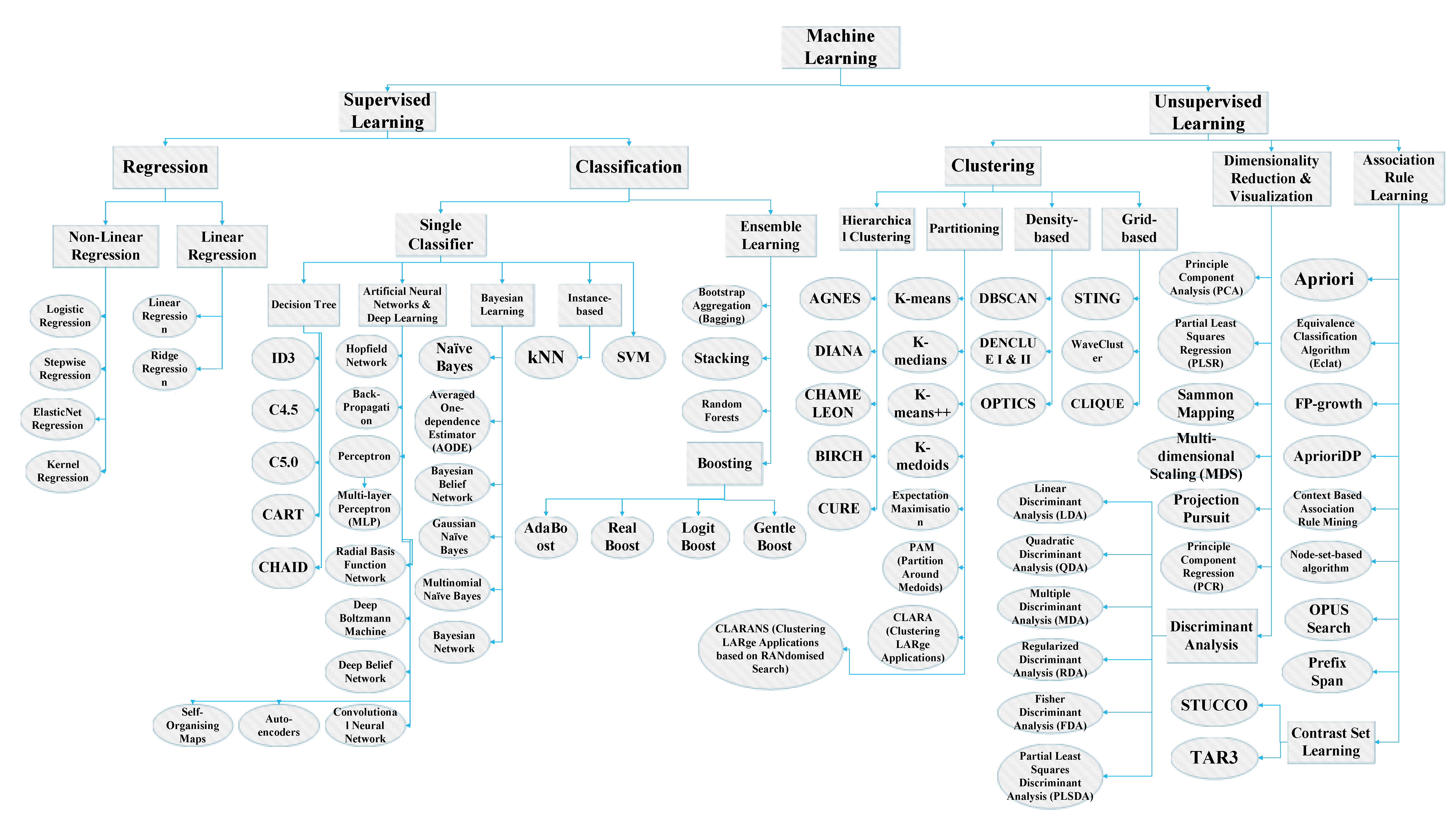}
 \caption{A taxonomy of Machine Learning algorithms for summarizing the large graphs~\cite{iSheets}.}
 \label{fig:iML}
 \end{figure}
\end{landscape}

In process analysis context, another common function applied on ER graphs is \emph{Process Discovery}. In particular, process mining techniques and tools offer a wide range of algorithms for discovering knowledge from process execution data.
In BP-SPARQL, we have identified many useful machine learning algorithms~\cite{iSheets} and provide an external algorithm controller to enable summarizing large graphs, by extracting complex data structures such as timeseries, hierarchies, patterns and subgraphs. 
Fig.~\ref{fig:iML}, illustrates a taxonomy of Machine Learning algorithms for summarizing large ER graphs.
We provide an extensible interface to support external graph reachability algorithms (such as~\cite{yu2010graph,BPM11}: Transitive Closure, GRIPP, Tree Cover, Chain Cover, Path-Tree Cover, and Shortest-Path) to discover set of related paths and store them in path nodes.

\begin{definition}\label{definition:processModel} \emph{
(Process Model) A process model $PM=(PG,F_{pd})$ is a summarized representation of a process instances graph $PG$ obtained by applying a process discovery algorithm $F_{pd}$. 
}\end{definition}

To discover process models, a query can be applied on a previously constructed process instances graph (stored in a folder node). For example, a process instances graph, as a result of a correlation condition, may partition a subset of the events in the graph into instances of a process.

\section{Querying Big Process Data}
\label{queryFT}

In this section, we present a summary of our previous work~\cite{BPM11,caise13,POLAP,ProcessAtlas} on querying process data.
We present the BP-SPARQL (\underline{B}usiness \underline{P}rocess \underline{SPARQL}), an extension of SPARQL\footnote{Among languages for querying graphs, SPARQL \cite{SPARQL} is an official W3C standard based on a powerful graph matching mechanism.}
for analyzing business process execution data.
BP-SPARQL enables the modeling, storage and querying of process data using Hadoop MapReduce framework.
In BP-SPARQL, we use folder and path nodes as first class entities that can be defined at several levels of abstraction and queried.
BP-SPARQL provides a data mapping layer for the automatic translation of SPARQL queries into MapReduce operations.
To additionally account for the subjectiveness of process data analysis, as well as to enable process analysts to analyze business data in an easy way, we support various querying needs over process data.
Next, we present various types of queries supported by BP-SPARQL including: Entity Level (artifacts, events and activities), Summarization (OLAP Style, Group Style and Partition Style), Relationship (Regular-Expression, Path Condition and Path Node), Metadata (Time and Provenance) and User-Defined Queries.

\subsection{Entity Level Queries}

We support the use of SPARQL to query entities and their attributes. 
We introduce the \emph{entity} statement which enables process analysts to extract information about process related entities such as business artifacts, people, and activities in an easy way. This statement has the following syntax:

\scriptsize
\begin{Verbatim} [frame=single]
entity [entity-type] \[attribute-name] =/!=/>/>=/</<= [value]
\end{Verbatim}
\normalsize

In this statement, `entity' is a reserved word,
`entity-type' is the type of process related entities, such as artifact and people,
and the `value' represents the value of the entity. The value should be quoted. The `\textbackslash' character represents the filter to be applied to entity attributes. The entity statement supports the conditional AND and OR operators. Parenthesis can be used in complex filters. The result of this query is an entity or a set of entities satisfying the condition.
The entity statement automatically translates to a SPARQL query. Details about this translation process can be found in~\cite{BPM11}.

\begin{example}\label{example:elq} \emph{Considering the motivating scenario, Tim (a process analyst) is interested in finding home-loan documents submitted to `Sydney' branches.}
\end{example}

\scriptsize
\begin{Verbatim} [frame=single]
entity artifact \category=`home-loan' AND \submission-branch=`Sydney'
\end{Verbatim}
\normalsize

In this example, `artifact' is the type of the entity to be filtered, ``$\backslash$category= `home-loan'" filters the category of the output entities to home-loan documents, and ``$\backslash$submission-branch=`Sydney'" filters the output to entities whose `submission branch' attribute is set to `Sydney'. More complex filters can be applied in this query. For example, if Tim is interested to find artifacts submitted in December 2017, he needs to add ``($\backslash$submission-date $\ge$ `01-12-2017' AND $\backslash$submission-date $\le$ `31-12-2017')" condition to the above query.

\subsection{Summarization Queries}

BP-SPARQL supports three types of summarization queries: OLAP Style, Group Style and Partition Style queries.

\emph{OLAP Style Queries.}
BP-SPARQL supports scalable graph-based OLAP analytics over process execution data~\cite{POLAP,DBLP:conf/wise/BeheshtiBNA12}.
The goal is to facilitate the analytics over the ER graph through summarizing the process graph and providing multiple views at different granularities.
To achieve this goal, we present a process OLAP (P-OLAP) model and define OLAP specific abstractions in process context such as process cubes, dimensions, and cells.
For example, analytics queries can be used to partition the ER graph in the example scenario into sets of related actors collaborating on (specific) loan applications. To achieve this, set of dimensions~\cite{POLAP} coming from the attributes of customer, loan documents, actors and the relationship among them should be analyzed. 

\emph{Group Style Queries.}
To summarize the process data,
BP-SPARQL extensively supports multiple information needs with one data structure (ER graph) and one function (machine learning algorithms presented in Fig.~\ref{fig:iML}). This capability enables analysts summarizing the large process graph, by extracting complex data structures such as timeseries, hierarchies, patterns and subgraphs. 

\emph{Partition Style Queries.}
%
A correlation condition query can be used to partition (a subset of) the entities in the graph into related instances.
Such partition style queries, enable the analyst to divide the process data into partitions that can be stored in folder nodes and accessed separately.
For example, it can be used to partition the events in the process graph into a set of process instances.
This statement has the following syntax:

\scriptsize
\begin{Verbatim} [frame=single]
correlation [Correlation-Condition]
\end{Verbatim}
\normalsize

In this statement, `correlation' is a reserved word and `Correlation-Condition' is the condition to be defined.
The result of this query is a collection of related entities satisfying the condition.
The correlation statement is automatically translated to a BP-SPARQL query. Details about this translation process can be found in~\cite{BPM11}.

\begin{example}\label{example:Correlation} \emph{Tim is interested in partitioning the graph in the example scenario into a set of related entities having the same type (e.g., customer, actors, and document). The correlation condition $\psi (node_x ,node_y) : node_x.type = node_y.type$ can be defined over the attribute $type$ of two node entities $node_x$ and $node_y$. 
This predicate is true when $node_x$ and $node_y$ have the same type and false otherwise.
Related node entities will be stored in folders, where each folder conforms to an entity type described by a set of attributes.}
\end{example}

A correlation condition can be assigned to a folder node to store the result of the query.
Also, timed folders~\cite{caise13} can be used to document the evolution of the folder over time. A monitoring code snippet can be assigned to a folder, e.g., to execute the correlation condition query over time or execute the query in case of triggers. In this case, new entities can be added to timed folders over time.


\subsection{Regular-Expression Queries}
\label{relationship-queries}

Regular-Expression queries can be used to discover transitive relationships between two entities in the ER graph.
In order to discover transitive relationships among entities,
BP-SPARQL supports regular language reachability algorithms~\cite{yu2010graph,BPM11} over ER graphs. The result of such a query is stored in a path node.
In particular, BP-SPARQL is designed to be customizable by process analysts who can codify their knowledge into \emph{regular expressions} that describe paths through the nodes and edges in the ER graph.

A path through the graph recognized by a regular expression would be useful if, by computing the path and its endpoint nodes, it answers a \emph{question} posed by a process analyst.
In this context, regular expressions are written by domain experts and will be executed over the ER graph. BP-SPARQL takes care of the processing and optimization that is needed for efficient crawling, analyzing and querying of the graph. We introduce the \emph{relationship} statement which enables process analysts to discover useful patterns among process related entities. 
This statement has the following syntax:

\scriptsize
\begin{Verbatim} [frame=single]
relationship [Regular-Expression]
\end{Verbatim}
\normalsize

In this statement, `relationship' is a reserved word and `Regular-Expression' is a parameter. 
This statement will be automatically translated into a path query in BP-SPARQL. 
Details about this translation process can be found in~\cite{BPM11}.
The following examples illustrate how a domain expert can use regular expressions to discover transitive relationships between business artifatcs, people, and activities:

\begin{example}\label{example:path1}
\emph{Find bank staff that is working on Adam's home loan document.\\
\scriptsize
\textbf{Regular Expression:} \\
\textcolor{black}{Adam (edge node)* assigned-to Staff}\\
\textbf{Example discovered path:}\\
\textcolor{black}{Adam $\xrightarrow{(submitted)}$ document $\xrightarrow{(part-of)}$ work-item $\xrightarrow{(assigned-to)}$ Staff}
\normalsize
}
\end{example}

\begin{example}\label{example:path2}
\emph{Find manager that approved Adam's loan report.\\
\scriptsize
\textbf{Regular Expression:} \\
\textcolor{black}{Adam (edge node)+ approved-by Manager}\\
\textbf{Example discovered path:}\\
\textcolor{black}{Adam $\xrightarrow{(submitted)}$ document $\xrightarrow{(part-of)}$ work-item $\xrightarrow{(assigned-to)}$ staff $\xrightarrow{(created)}$ report $\xrightarrow{(approved-by)}$ Manager}
}
\normalsize
\end{example}

\begin{example}\label{example:path3}
\emph{Find artifacts related to Adam.\\
\scriptsize
\textbf{Regular Expression :} \\
\textcolor{black}{Adam (edge node)* edge Artifact}\\
\textbf{Example discovered paths:}\\
\textcolor{black}{Adam $\xrightarrow{(submitted)}$ Home-Loan-Document}\\
\textcolor{black}{Adam $\xrightarrow{(submitted)}$ document $\xrightarrow{(part-of)}$ work-item $\xrightarrow{(assigned-to)}$ Tim $\xrightarrow{(created)}$ Home-Loan-Report}
}
\normalsize
\end{example}

In these examples, regular expressions are used to discover sets of paths in the process graph.

\subsubsection{Path Condition Queries}

are similar to relationship queries, however, they are able to store the query results in folder nodes.
In particular,
a \emph{path condition}~\cite{POLAP} can be used to group related entities in the ER graph based on set of dimensions coming from the attributes of network structures: we need to apply conditions not only on graph entities but also on the relationships among them.
A path condition $\phi$ is defined as a binary predicate on the attributes of a path that allows to identify whether two or more entities (in a given ER graph) are potentially related through that path.

For example, Tim is interested in finding a set of related actors (e.g. home loan employees) working on Adam's home loan application.
Therefore, Tim is interested in creating a folder node for a set of related actors and then adding related actors to this folder if: there exists a specific path between the customer and an actor.
The path condition $\phi (node_{start},node_{end},RE)$ can be defined on the existence of the path codified by the regular expression RE:[\emph{Adam (edge node)* assigned-to STAFF}] between starting node, $node_{start}$, and ending node, $node_{end}$. This predicate is \emph{true} if the path exists and \emph{false} otherwise. 
A path condition can be assigned to a folder node to store the result of the query (set of entities).
Details about this type of queries can be found in~\cite{POLAP}.

\subsubsection{Path Node Queries}
\label{path-Queries}

There are situations where process analysts are interested in storing the discovered paths as a result of a relationship query. 
This will enable to store a set of related patterns in a path node and use them as an input for further analytics tasks. Notice that, results of relationship queries may be different over time, as new nodes and relationships can be added over time.
Details about this type of queries can be found in~\cite{POLAP}.

\subsection{Metadata Queries}
\label{Metadata-Queries}

In a process execution path, huge amount of process related meta-data, such as versioning (what are the various versions of an artifact during its lifecycle~\cite{DBLP:conf/adc/MaamarSBB15},
and how are they related), provenance (what manipulations were performed on the artifact to get it to this version), security (who has access to the artifact over time), and privacy (what actions were performed to protect or release artifact information) can be recorded. These metadata can be used to imbue the process data with additional semantics. 

In~\cite{caise13}, we formalized metadata to be collected while an activity is taking place, including:
\begin{itemize}
  \item \emph{When}, to indicate the timestamp in which the activity has occurred;
  \item \emph{Who}, to indicate an actor that enables, facilitates, controls, or affects the activity execution;
  \item \emph{Where}, to indicated the organization/department the activity happened;
  \item \emph{Which}, to indicate the system which hosts the activity; and
  \item \emph{Why}, to indicate the purpose of the activity, e.g., fulfilment of a specific phase or experiment;
\end{itemize}

We highlighted that discovering paths through ER graphs forms the basis of many metadata queries.
Next, we present simple queries to discover evolution (how the business artifact evolved over time?), derivation (what are the ancestors of the business artifact?), and timeseries (what are the snapshots of the business artifact over time?) of business related artifacts. The query template statement has the following syntax:

\scriptsize
\begin{Verbatim} [frame=single]
metadata evolutionOf/derivationOf/timeseriesOf  [artifact-name]
filter [who, where, which, when, ...]
\end{Verbatim}
\normalsize

This statement can be used for discovering evolution of artifacts (using evolutionOf construct),
derivation of artifacts (using derivationOf construct), and timeseries of artifacts/actors
(using timeseriesOf construct). The `filter' statement restricts the result to those activities for which the filter expression evaluates to true. Variables, such as artifact, type (e.g., lifecycle or archiving), action (e.g., creation, use, or storage), actor, and location (e.g., organization) are defined as filters.

\begin{figure*}[t]
\centering
  \includegraphics[scale=0.71]{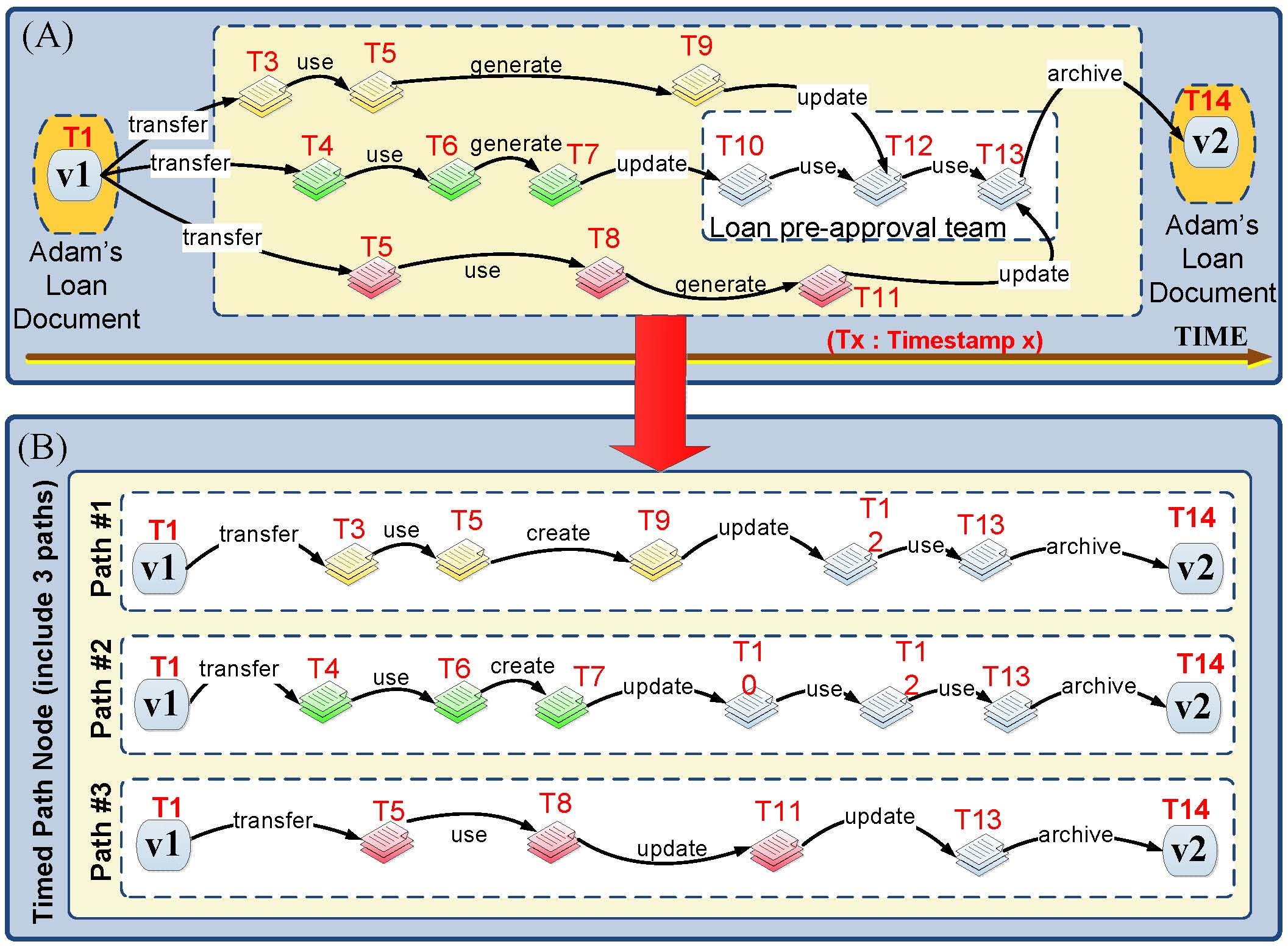}
  \caption{Example business process executions.}\label{evolution-example}
\end{figure*}

\begin{example}\label{example:evolution} \emph{For querying the evolution of an entity $En$, all activity paths on top of $En$ ancestors should be discovered. For example, considering the motivating scenario, Tim is interested to see how version $v_2$ of Adam's loan document evolved from version $v_1$.}
\end{example}

Following is a sample query template for this example.

\scriptsize
\begin{Verbatim} [frame=single]
evolutionOf Adam_loan_document_v2
\end{Verbatim}
\normalsize

Fig.~\ref{evolution-example}-A illustrates the result of this query. Fig.~\ref{evolution-example}-B, illustrates how a set of paths between the two versions can be stored in a sample path node. In particular, three paths are recognized, assigned a unique identifies (e.g., $path\#1$) and stored under a path node name. Further user-defined queries can be applied to the path nodes for follow on analysis.
Additional filters can be added to the above query.
For example, if Tim is interested to see the activities which involved creating a new artifact, he can use the following query template:

\scriptsize
\begin{Verbatim} [frame=single]
evolutionOf Adam_loan_document_v2 \what=`lifecycle' \how=`create'
\end{Verbatim}
\normalsize

As a result, Tim will only see paths $path\#1$ and $path\#2$, in Fig.~\ref{evolution-example}-B.
More examples can be found in~\cite{caise13}.

\subsection{User-Defined Queries}

Besides the above mentioned query templates, a process analyst may apply user defined queries. 
In this case, the analyst should be familiar with SPARQL syntax. In particular, a basic SPARQL query has the following form:

\scriptsize
\begin{Verbatim} [frame=single]
select ?variable1 ?variable2 ...
where {pattern1. pattern2. #Other-Patterns}
\end{Verbatim}
\normalsize

Each pattern consists of \emph{subject}, \emph{predicate} and \emph{object}, and each of these can be either a variable or a literal.
The query specifies the known literals and leaves the unknown as variables.
To answer a query one needs to find all possible variable bindings that satisfy the given patterns.
It is possible to use the `@' symbol for representing attribute edges and distinguishing them from the relationship edges between graph nodes.
As an example, considering the motivating scenario, Tim may be interested in monitoring the conversations (e.g., messages) between the loan approval team. For example, Tim may be interested in retrieving a list of messages that have the same value on `requestsize' and `responsesize' attributes and the values for their timestamps falls between $t_1$ and $t_2$. Following is the SPARQL query for this example:

\scriptsize
\begin{Verbatim} [frame=single]
1:  select ?m
2:  where{
3:    ?m @type message.
4:    ?m @requestsize ?x.
5:    ?m @responsesize ?y.
6:    ?m @timestamp ?t.
7:    FILTER (?x=?y && ?t > t1 && ?t < t2). }
\end{Verbatim}
\normalsize

In this query, variable $?m$ represents a message in the graph. Variables $?x$, $?y$, and $?t$ represent the value of the attribute `requestsize' (line 4), `responsesize' (line 5), and `timestamp' (line 6), respectively. The \emph{FILTER} statement restricts the result to those messages for which the filter expression evaluates to \emph{true}.

\section{Scalable Analysis Using MapReduce}
\label{MapReduceSupport}

The popular MapReduce~\cite{MapReduce} scalable data processing framework, and its opensource realization Hadoop~\cite{Hadoop}, offer a scalable data
flow programming model that appeals to many users.
In MapReduce frameworks, computations are specified via two user-defined functions: a mapper that takes key-value pairs as input and produces key-value pairs as output, and a reducer that consumes those key-value pairs (generated in the mapper phase) and aggregates data based on individual keys.
In practice, the extreme simplicity of the MapReduce programming model leads to several problems. For example, it does not directly support complex N-step data flows which often arise in practice. To address this problem, Apache Pig~\cite{olston2008pig} system offers compassable high-level data
manipulation constructs in the spirit of SQL, while at the same time retaining the properties of MapReduce systems, 
makes them attractive for certain users, data types, and workloads.

Pig's language layer consists of a textual language called PigLatin which supports ease of programming,
optimization opportunities, and extensibility. It is possible to write a single script in PigLatin that
is automatically parallelized and distributed across a Hadoop~cluster.
A script in Pig often follows the $input-process-output$ (IPO) model:
(i)~Input: data is read from the Hadoop Distributed File System (HDFS);
(ii)~Process: a number of operations (e.g., LOAD, SPLIT, JOIN, FILTER, GROUP, and STORE) are performed on the data; and
(iii)~Output: the resulting relation is written back to the file~system.

\begin{figure*}[t]
\centering
  \includegraphics[scale=0.56]{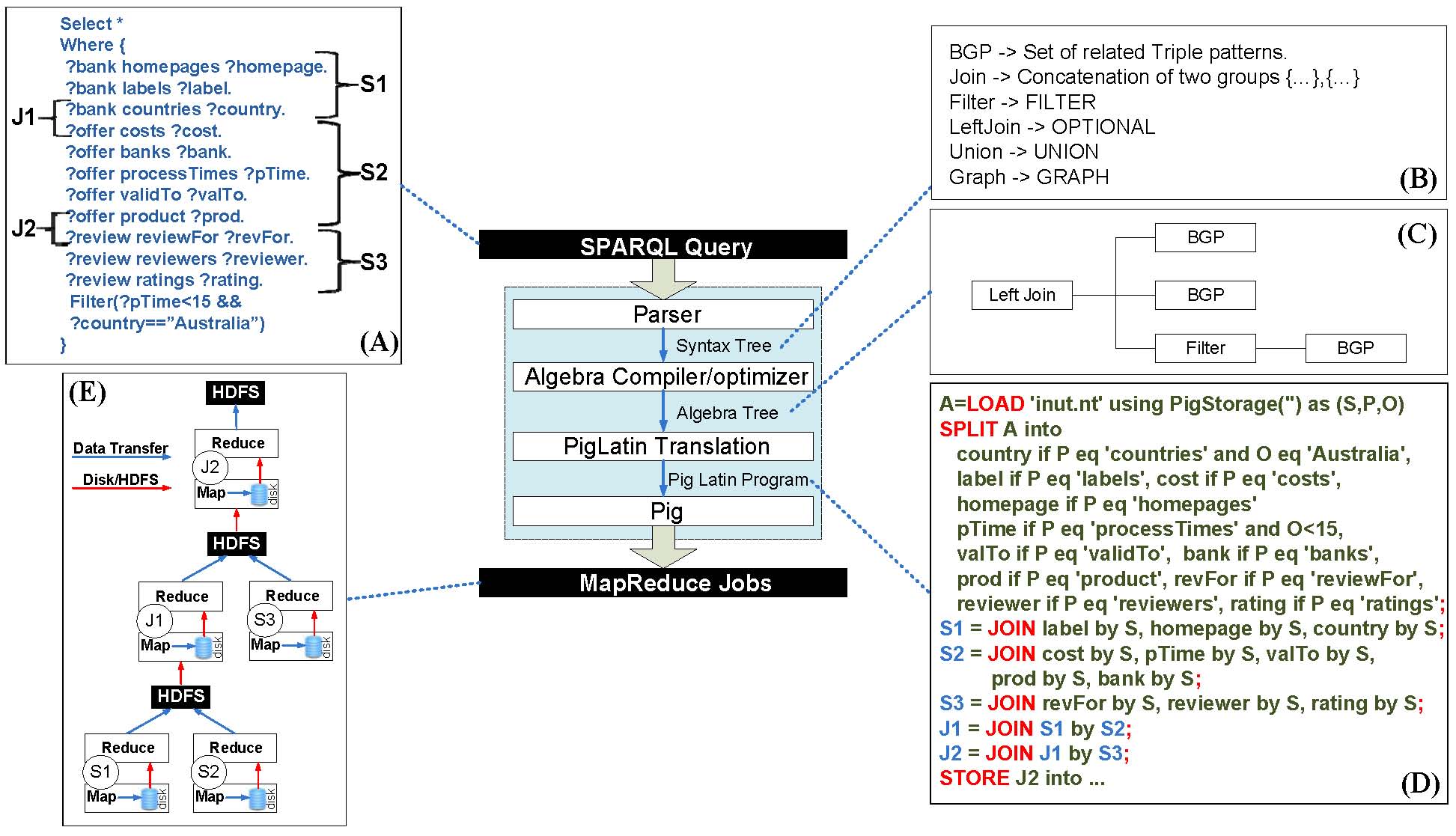}
  \caption{An example BP-SPARQL query~\cite{POLAP,ProcessAtlas} (A) the SPARQL syntax tree (B); and Algebra tree (C), translated PigLatin (D) and its MapReduce execution flow (E).}\label{MRenabled}
\end{figure*}

Retrieving related graphs 
containing a graph query from a large RDBMS graph database is a key performance issue, where a primary challenge in computing the answers of such graph queries is that pairwise comparisons of graphs are usually hard problems.
We use Hadoop~\cite{Hadoop} data processing platform to store and retrieve graphs in Hadoop file system, and to support cost-effective and scalable processing of graphs. We use Apache Pig, a high-level procedural language on top of MapReduce, for querying large graphs stored in Hadoop file system.
We use the algebra proposed in~\cite{PigSPARQL} for mapping SPARQL queries to PigLatin programs and consequently generating MapReduce operations.
To capture the storage model in Pig, an input graph needs to be `split' into property-based partitions using PigLatin's SPLIT command. Then, the star-structured joins are achieved using the m-way JOIN operator, and chain joins are executed using the binary JOIN operator.

PigLatin queries are compiled into a sequence of MapReduce operations that run over Hadoop. The Hadoop scheduling supports partition parallelism such that in every stage,  one operator is running on a different partition of data.
In particular, the logical plan for Pig queries can be described as follows: (i)~Load the input dataset using the LOAD operator; (ii)~Create vertical partitioned relations using the SPLIT operator; (iii)~Join partitioned relations based on join conditions: In SPARQL, join conditions are implied by repeated occurrences of variables in different triple patterns. Consequently, for each star join, the join will be computed in a single
MapReduce cycle; and (iv)~Store the result on disk using the STORE operator.

For example, consider the following query in the context of the motivating scenario: ``give the name of Australian banks which offer loan products (e.g., home loan, business loan, variable/fix rates, etc.) within 15 days, along with the review details for these products". Fig.~\ref{MRenabled} illustrates the corresponding BP-SPARQL query and MapReduce execution flow. As illustrated, the query can be factorized into three main sections: (three star-join
structures (S1, S2, S3) describing resources of type Vendor, Offer, and Review respectively, two chain-join pattern (J1, J2) combining the star patterns, and the filter processing. In particular, such queries can be considered equivalent to the select-project-join construct in SQL, where each MapReduce cycle may involve communication and I/O costs due to the data transfer between the mapper and the reducer.

\section{Implementation}
\label{imp}

We have implemented a research prototype of BP-SPARQL platform for organizing, indexing, and querying process data.
The prototype supports two types of storage back-end: Relational Database System and Hadoop File System. We use Apache-Pig (pig.apache.org), a high-level procedural programming language on top of Hadoop for querying large graphs.
The main components of the query engine include:
(i)~\emph{Data Mapping Layer}: This layer is responsible for creating data element mappings between semantic web technology 
and physical storage layer, i.e., relational database schema and Hadoop File System;
(ii)~\emph{Time-aware Controller}: RDF databases are not static and changes may apply to graph entities (i.e., nodes, edges, and folder/path nodes) over time. Time-aware controller is responsible for data changes, data manipulation, and incremental graph loading;
(iii)~\emph{Query Mapping Layer}: This layer is responsible for BP-SPARQL queries translation and processing;
(iv)~\emph{Regular Expression Processor}: We developed a regular expression processor which supports optional elements, loops, alternation, and groupings.
(v)~\emph{External Algorithm/Tool Controller}: This component is responsible for supporting external graph reachability/mining algorithms; and
(vi)~\emph{GOLAP Controller}: This component is responsible for partitioning graphs and evaluation of OLAP operations independently for each partition, providing a natural parallelization of execution.

\begin{figure}
  \centering
  \includegraphics[width=1.0\textwidth]{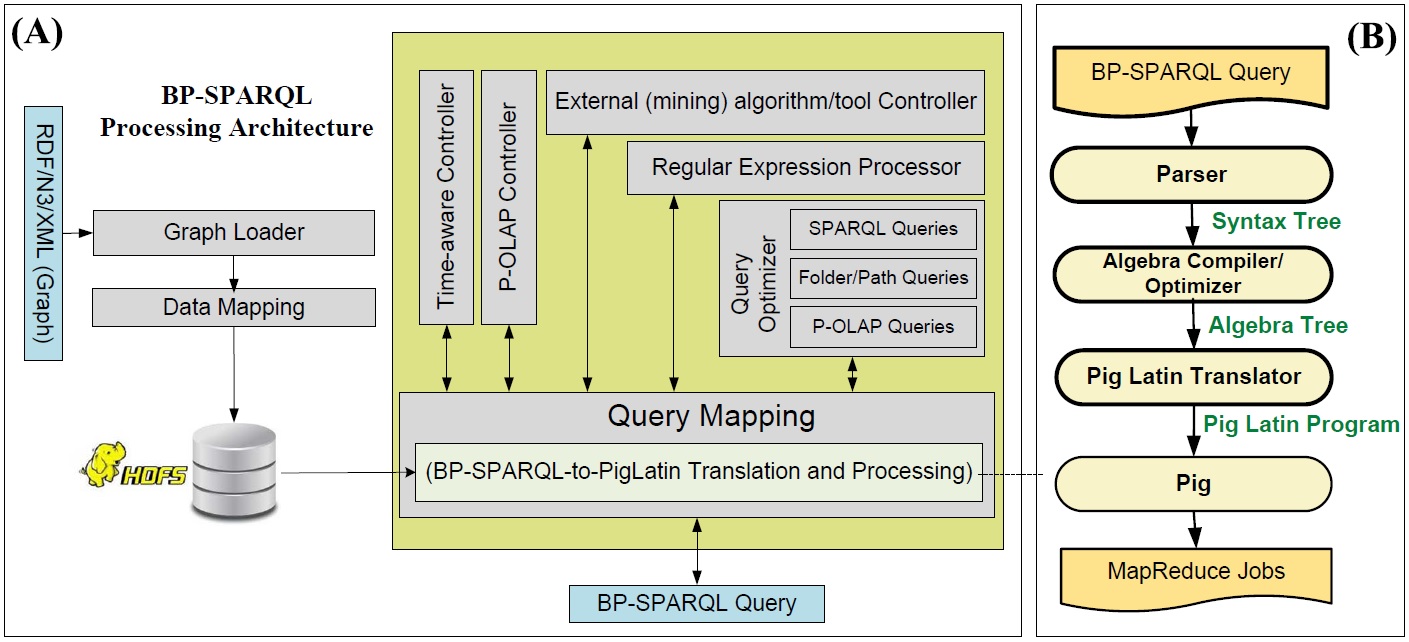}\\
  \caption{BP-SPARQL graph processing architecture (A) and the modular translation process for mapping SPARQL to PigLatin (B)~\cite{POLAP}.} \label{graphArchitecture}
\end{figure}

Details on the implementation and evaluation can be found in our previous work~\cite{BPM11,caise13,POLAP,ProcessAtlas}.
Figure~\ref{graphArchitecture}(A) illustrates BP-SPARQL graph processing architecture.

\section{Process Querying Framework}
\label{RelWork}

Polyvyanyy et al.~\cite{DBLP:journals/dss/PolyvyanyyOBA17} proposed
the framework for developing process querying methods.
%
Complimentary to the active components proposed in this framework, 
BP-SPARQL focused on facilitating the querying and analysis of data-driven processes and knowledge intensive processes.
Such processes include a set of coordinated tasks and activities, controlled by knowledge workers to achieve a business objective or goal.
Examples include police investigation processes and government processes, such as in immigration, health and educations departments.
%
Facilitating the querying and analysis of such processes is important, as the continuous improvement in connectivity, storage and data processing capabilities allow access to a data deluge from sensors, social-media, news, user-generated, government and private data sources.
Accordingly, business processes become inseparable from data. Examples of process data include: data from the execution of business processes, documentation and description of processes, process models, process variants, artifacts related to business processes, and data generated or exchanged during process execution.
In the following, we discuss some of the related works in organizing, querying and analyzing process data.

\textbf{From Data Lakes to Knowledge Lakes.}
With data science continuing to emerge as a powerful differentiator across industries, almost every organization is now focused on understanding their business and transforming data into actionable insights.
The notion of a Data Lake~\cite{DBLP:conf/cikm/BeheshtiBNCXZ17} has been coined to address this challenge and to convey the concept of a centralized repository containing limitless amounts of raw (or minimally curated) data stored in various data islands. The rationale behind a Data Lake is to store raw data and let the data analyst decide how to cook/curate them later. While Data Lakes do a great job in organizing Big Data and providing answers on known questions, the main challenges are to understand the potentially interconnected data stored in various data islands and to prepare them for analytics.

The notion of Knowledge Lake~\cite{DBLP:journals/pvldb/BeheshtiBNT18}, i.e., a contextualized Data Lake, introduced to automatically transform the raw (process) data into contextualized data and knowledge.
The term Knowledge here refers to a set of facts, information, and insights extracted from the raw data using data curation techniques, such as extraction, linking, summarization, annotation, enrichment, classification and more~\cite{DataSynapse,intelKG}. In particular, a Knowledge Lake is a centralized repository containing virtually inexhaustible amount of data and contextualized data that is readily made available anytime to anyone authorized to perform analytical activities. Knowledge Lakes provide the foundation for Big Data analytics by automatically curating the raw data in Data Lakes and preparing them for deriving insights.

\textbf{Process Graph Modeling.}
Graphs are essential modeling and analytical objects for representing information networks.
Several graph querying techniques~\cite{GraphBook} such as pattern match query, reachability query, shortest path query, and subgraph search, have been proposed for querying and analyzing graphs.
These methods rely on constructing some indices to prune the search space of each vertex to reduce the whole search space.
In~\cite{GraphBook}, authors discuss a number of data models and query languages for graph data. Many of these models use RDF~\cite{SPARQL} (Resource Description Framework), an official W3C recommendation for semantic Web data models, to model graphs and use SPARQL, an official W3C recommendation for querying RDF graphs~\cite{SPARQL}. SPARQL queries are pattern matching queries on triples that constitute an RDF data graph, where RDF is a data model for schema-free structured information.
Several research efforts have been proposed to address efficient and scalable management of RDF data. 
SPARQL \cite{SPARQL} is a declarative query language, an W3C standard, for querying and extracting information from directed labeled RDF graphs. SPARQL supports queries consisting of triple patterns, conjunctions, disjunctions, and other optional patterns. However, there is no support for querying grouped entities. Paths are not first class objects in SPARQL~\cite{SPARQL}. PSPARQL \cite{PSPARQL} extends SPARQL with regular expressions patterns allowing path queries. SPARQLeR \cite{SPARQLeR} is an extension of SPARQL designed for finding semantic associations (and path patterns) in RDF graphs.
In BP-SPARQL, we support folder and path nodes as first class entities that can be defined at several levels of abstractions and queried. 

\textbf{Knowledge-Intensive Processes.}
Case-managed processes are primarily referred to as semistructured processes, since they often require the ongoing intervention of skilled
and knowledgeable workers. Such Knowledge-Intensive Processes involve operations that rely on professional knowledge.
For these reasons, it is considered that human
knowledge workers are responsible to drive the process, which cannot otherwise
be automated as in workflow systems~\cite{ProcessAtlas}.
Knowledge-intensive processes often involve the collection and presentation of a diverse set of artifacts and human activities around artifacts. This emphasizes the artifact-centric nature of such processes.
Many approaches~\cite{artifactCenter2,BhattacharyaGHLS07,DBLP:journals/tmis/SunSY16} use business artifacts that combine data and processes in a holistic manner and as the basic building block. A line of work~\cite{artifactCenter2} used a variant of finite state machines to specify lifecycles. A theoretical work~\cite{BhattacharyaGHLS07} explored declarative approaches to specifying the artifact lifecycles following an event oriented style.
Another line of work in this category, focused on querying artifact-centric processes~\cite{documentDriven}.

\textbf{Process Data Analytics.}
In our recent book~\cite{AminBook}, we provided an overview of the state-of-the-art in the area of business process management in general and process data analytics in particular. This book provides defrayals on: (i) technologies, applications and practices used to provide process analytics from querying to analyzing process data; (ii) a wide spectrum of business process paradigms that have been presented in the literature from structured to unstructured processes; (iii) the state-of-the-art technologies and the concepts, abstractions and methods in structured and unstructured BPM including activity-based, rule-based, artefact-based, and case-based processes; and (iv) the emerging trend in the business process management area such as: process spaces, big-data for processes, crowdsourcing, social BPM, and process management in the cloud.
BPM in the Cloud solutions offer visibility and management of business processes, low start up costs and fast return on investment.
Crowdsourcing can help organizations to increase productivity by discovering and exploiting informal knowledge and relationships to improve activity execution. Crowdsourcing can also enable the socialBPM to assign an activity to a broader set of performers or to find appropriate contributors for its execution.
Social BPMs inevitably require advanced crowd-management capabilities in future social computing platforms.

\section{Conclusion}
\label{Conclusion}

The continuous demand for the business process improvement and excellence has prompted the need for business process analysis. 
Recently, business world is getting increasingly dynamic as various technologies such as Internet and email have made dynamic processes more prevalent.
In this chapter, we focused on the problem of explorative querying and understanding of business processes data. Our study shows that only part of interactions related to the process executions are covered by process-aware systems as business processes are realized over a mix of workflows, IT systems, Web services and direct collaborations of people. In order to fulfill the requirements, we proposed a framework for organizing, indexing, and querying ad-hoc process data. In this framework, we proposed novel abstractions for summarizing process data, and a language, BP-SPARQL, for the explorative querying and understanding of BP execution from various user perspectives.

In future work, we will employ interactive graph exploration and visualization techniques (e.g., storytelling systems~\cite{iStory,ConceptMap,samiSum1}) to facilitate the use of BP-SPARQL through visual query interface.
We will focus on significant scientific advancement in understanding the practical problems of leveraging cognitive assistants in facilitating knowledge-intensive processes~\cite{DBLP:conf/birthday/BarukhZBBBCYSS21,icop,iRecruit}. We aim to develop novel techniques in linking cognitive science and data science to enable personalizing of process automation~\cite{CognitiveRecom,personality2vec,icopCig}. This will also plan to facilitate the analysis of personality aspects in personalized recommendations~\cite{recom1,recom2,recom3}.
As an ongoing and future work, we are developing an intelligent business rules engine to support adaptive rule adaptation in unstructured and dynamic environments~\cite{rule1,rule2,rule3}.


\bibliographystyle{plain}
\bibliography{Biblio}

\end{document}